\documentclass[a4paper]{article}

\usepackage{graphicx}
\usepackage{amsfonts}
\usepackage{amsmath}
\usepackage{amsbsy}
\usepackage{subfigure}

\begin{document}

\title{Dynamic Non-Null Magnetic Reconnection in Three Dimensions - II. Composite Solutions}

\author{ A.L. Wilmot-Smith \\  { \small \it Division of Mathematics, University of Dundee, Old Hawkhill Road, Dundee DD1 4HN, UK} \\
G. Hornig \\  {\small \it Division of Mathematics, University of Dundee,  Old Hawkhill Road, Dundee DD1 4HN, UK} \\
and \\  E.R.  Priest \\ {\small \it School of Mathematics and Statistics, University of St Andrews, St Andrews KY16 9SS, UK}}

\maketitle

\begin{abstract}
\noindent
In this series of papers we examine magnetic reconnection in a domain where the magnetic field does not vanish 
and the non-ideal region is localised in space.  In a previous paper we presented a technique for obtaining analytical 
solutions to the stationary resistive MHD equations in such a situation and examined specific examples of non-ideal 
reconnective solutions.  Here we further develop the model,  noting that certain ideal solutions may be superimposed 
onto the fundamental non-ideal solutions and examining the effect of imposing various such flows.
 Significant implications are found for the evolution of magnetic flux in the reconnection process.
It is shown that, in contrast to the two-dimensional case, in three-dimensions there is a very wide variety 
of physically different steady reconnection solutions.
\end{abstract}

\section{Introduction}

Magnetic reconnection is a fundamental process in astrophysical, space and laboratory plasmas. 
The first models of the process were two-dimensional (2D) and stationary; much of our present understanding of 
the subject (for a review see, for example, Priest \& Forbes, 2000) relies on these models.
However, astrophysical plasmas systems are typically three-dimensional (3D) with any non-ideal regions localized 
in each dimension.  Models incorporating these features have demonstrated important differences 
between the 3D and 2D cases (Hornig \& Priest, 2003).   In this paper we focus on the basic properties of 3D steady magnetic  
reconnection in a resistive MHD plasma with the aim of exploring these new features.

In 2D, a change in the magnetic connectivity of plasma elements occurs at an X-type magnetic null-point 
that is located in a non-ideal region of an otherwise ideal plasma.
Through analytical modelling, a range of steady-state 2D  resistive MHD reconnection theories have been 
developed 
(e.g., Petschek, 1964; Vasyliunas, 1975; Priest \& Forbes, 1986;  Priest \& Lee, 1990;  Craig \& Henton, 1995).
Expansion procedures are one of the techniques that have proved invaluable in the development of 
these models,  with the small parameter commonly taken as the Alfv{\'e}n Mach number
of the flow.   The early paper by Petsheck (1964) used such an expansion and, subsequently, 
Priest \& Forbes (1986) applied the method to develop  a family of almost-uniform regimes in which the nature 
of the inflow velocity determining  the particular regime and with Petschek's mechanism being a special case.  
However, all 2D models have a number of elements in common;
a stagnation flow that crosses the separatrices of the field brings magnetic flux in toward and, subsequently, 
away from, the non-ideal region and, with the electric field being uniform, the rate of reconnection is governed 
by the solution in the  ideal region.

In 3D, reconnection may also occur at null points, now in various forms depending on the direction of the current 
(Pontin \textit{et al.} 2004, 2005a), as well as along separators connecting null points (Lau \& Finn, 1990, 
Priest \& Titov, 1996) and at locations where the field does not vanish (Schindler \textit{et al.} 1988,   
Priest \& Forbes, 1992).    This {\it non-null reconnection}  requires only a localised non-ideal 
region together with a non-zero integrated electric field component parallel to the magnetic field  
(Hesse \& Schindler 1988). 
Each of these forms of 3D reconnection may look quite different, both from each other and from the
heavily analysed 2D case.

Regarding the specific case of non-null magnetic reconnection and a localised non-ideal region, 
the subject of the present study, 
 Hornig \& Priest (2003) presented a kinematic  stationary model of the process. Here the term {\it kinematic} 
 refers to the situation where the magnetic field is imposed and the equation of motion neglected.
The authors noticed that in 3D, under the kinematic
assumption,  Ohm's law can now be decomposed into non-ideal and ideal components:
\begin{eqnarray*}
\mathbf{E}_{non-ideal} + \mathbf{v}_{non-ideal} \times \mathbf{B} & = &  \eta \mathbf{j} \\
\mathbf{E}_{ideal} + \mathbf{v}_{ideal} \times \mathbf{B} & = & \mathbf{0}
\end{eqnarray*}
As a consequence of the localisation of the non-ideal region, the plasma flows in the non-ideal solution,
termed the  `particular solution' are rotational in nature,  with the rotation being in different senses 
above and  below the non-ideal region. An important feature of the particular solution is that it is spatially restricted 
with the rotational flows being found only within the field lines threading the non-ideal region.
Onto this particular solution can be imposed any solution to the ideal Ohm's law, the results 
being termed `composite solutions'.   Such solutions can ease the spatial confinement of the reconnection process, 
allowing its effect to be seen on a global scale.  However, the independence of the non-ideal and ideal
solutions suggests the characteristic coupling in 2D between large-scale flows and reconnection rate 
no longer holds in 3D.

In a previous paper (Wilmot-Smith, Hornig \& Priest, 2006, hereafter Paper I) we introduced an expansion scheme 
through which we may carry out a dynamic analysis of 3D non-null reconnection (i.e. including also the equation 
of motion)  and so address fundamental questions regarding both the apparent freedom in the 3D system
 (as suggested by Hornig \& Priest, 2003),  and the nature and diversity of non-null reconnection solutions.
In that introductory paper we developed the solution method and demonstrated how the system can be 
broken down into particular and composite solutions. We then proceeded to examine the case of particular 
solutions, again finding counter-rotational plasma flows confined to be within the flux-tube threading the non-ideal
region. However,  in the dynamic analysis the ideal and non-ideal components to the 3D Ohm's law are coupled 
via the equation of motion.
Although we may again impose an ideal solution onto the particular solution and create composite solutions, 
the coupling will restrict which ideal solutions may be taken.
We address this important question in Section~\ref{sec:composites} and find certain restrictions on the
choice of ideal flow. Accordingly we then proceed to examine an example of a more general ideal flow
in Section~\ref{sec:accel} where the precise division between particular and composite solutions no longer
holds but, nevertheless, a great deal of freedom within the 3D system is found.

\section{Model}

Using a method described in more detail in Paper I, we consider the dimensionless, stationary, 
incompressible, resistive MHD equations and  seek three-dimensional solutions for slow flows, 
(\begin{math} v \ll v_{A} \end{math}).    By assuming the inertial term in the momentum equation is small,
we may expand the dimensionless variables in terms of the  Alfv\'en Mach number of the flow, $M_{e} \ll 1$:
 \begin{displaymath} 
\mathbf{B} = \mathbf{B}_{0}+M_{e} \mathbf{B}_{1}+
            M_{e}^{2} \mathbf{B}_{2}+M_{e}^{3} \mathbf{B}_{3} + \cdots ,
\end{displaymath}
\begin{displaymath} 
\mathbf{v} = \mathbf{v}_{1}+M_{e} \mathbf{v}_{2}+
            M_{e}^{2} \mathbf{v}_{3} + \cdots ,
\end{displaymath}
 \begin{displaymath} 
\mathbf{j} = \mathbf{j}_{0} + M_{e} \mathbf{j}_{1}+
            M_{e}^{2} \mathbf{j}_{2}+M_{e}^{3} \mathbf{j}_{3} + \cdots ,
\end{displaymath}
 \begin{eqnarray*} 
\mathbf{E} &=& \mathbf{E}_{0}+M_{e} \mathbf{E}_{1}+
            M_{e}^{2} \mathbf{E}_{2} + \cdots \\
& = & -\nabla' \phi_{0} - M_{e}\nabla'\phi_{1} - M_{e}^{2}\nabla'\phi_{2} +\cdots,
\end{eqnarray*}
\begin{displaymath}
p =p_{0} + M_{e}p_{1} + M_{e}^{2}p_{2} + \cdots,
\end{displaymath}
where all the quantities 
\begin{math} \mathbf{B}_{i}, \ \mathbf{v}_{i}, \ \mathbf{j}_{i},
                          \ \mathbf{E}_{i}, \ \phi_{i}, \ p_{i} \end{math} 
are dimensionless.  

As in the analysis of Paper I we assume throughout that the lowest-order magnetic field, $\mathbf{B}_{0}$ 
is potential (and so $\mathbf{j}_{0}=\mathbf{0})$ and that $\mathbf{B}_{1}=\mathbf{0}$.
With these assumptions a comparison of the  zeroth- and first-order terms in the 
expansion of Ohm's law  yields ideal and non-ideal equations, respectively. These are given by
\begin{equation}
\label{eq:zerothohm}
\mathbf{E}_{0} + \mathbf{v}_{1} \times \mathbf{B}_{0} = \mathbf{0},
\end{equation}
\begin{equation}
\label{eq:firstohm}
 \mathbf{E}_{1} + \mathbf{v}_{2} \times \mathbf{B}_{0} = \hat{\eta} \mathbf{j}_{2}.
\end{equation}
In order to satisfy the continuity equation we are assuming for simplicity that
the plasma is incompressible. Accordingly,
$\nabla  \cdot \mathbf{v}_{1}=\nabla  \cdot \mathbf{v}_{2}=0$, together with
\eqref{eq:zerothohm} and \eqref{eq:firstohm}, are the only non-trivial 
equations of zeroth and first order.
The pressure terms $p_{0}$ and $p_{1}$ are both constants, and the main dynamic effects 
are described by the equation of motion at second-order:
\begin{equation}
\label{eq:secondmotion}
\left(\mathbf{v}_{1} \cdot \nabla\right)\mathbf{v}_{1} = -\nabla p_{2} + \mathbf{j}_{2} \times \mathbf{B}_{0}.
\end{equation}
This equation provides a coupling between \eqref{eq:zerothohm} and \eqref{eq:firstohm}.

Throughout the analysis we set
\begin{equation}
\label{eq:B0form}
\mathbf{B}_{0}=b_{0}(ky,kx,1),
\end{equation}
where $k>0$, so that the basic state is a current-free equilibrium representing the superposition
of an X-type null-point field in the $xy$-plane with a uniform $z$-component of the magnetic field. 
 The equations, 
\begin{math} \mathbf{X}(\mathbf{x}_{0},s) \end{math}, of the field line passing through the point 
\begin{math} \mathbf{x}_{0} \end{math}  are given by
\begin{equation}
\label{eq:fieldlines}
 X=x_{0}\cosh(b_{0}ks)+y_{0}\sinh(b_{0}ks),  \
 \ Y=y_{0}\cosh(b_{0}ks)+x_{0}\sinh(b_{0}ks), \ \
  Z=b_{0}s+z_{0},
  \end{equation}
with the inverse mapping \begin{math} \mathbf{X}_{0}(\mathbf{x},s) \end{math} given by
\begin{equation}
\label{eq:invfield}
  X_{0}=x\cosh(b_{0}ks)-y\sinh(b_{0}ks), \ \ 
 Y_{0}=y\cosh(b_{0}ks)-x\sinh(b_{0}ks), \ \
  Z_{0}=-b_{0}s+z,
  \end{equation}
where the parameter $s$ is related to the distance $l$ along a field line by $ds = dl/\vert B_{0} \vert$.

The lowest-order equations in the expansion procedure allow for a direct comparison with the particular
and composite solutions of Hornig \& Priest (2003), whose `particular solutions' are recovered in this
scheme by setting  \begin{math} \mathbf{E}_{0}=\mathbf{0} \end{math} and \begin{math} \mathbf{v}_{1}=\mathbf{0}. \end{math}
These solutions were described in some detail in paper I, where two examples of analytical solutions to 
\eqref{eq:firstohm} together with \eqref{eq:secondmotion} were given. In both cases the non-ideal term,
$\hat{\eta} \mathbf{j}_{2}$, is localised in all three-dimensions.  As a result of that localisation the reconnective
plasma flow, $\mathbf{v}_{2}$ is found to be confined to within the HFT and counter-rotational in nature
i.e.  rotating in opposite senses above and below the central $xy$-plane. Thus it is only the magnetic flux
linking the non-ideal region that is affected by the reconnection process.

In realistic situations we might expect a non-zero plasma velocity outside of the HFT.  One way to achieve this 
is to take a non-zero ideal plasma flow $\mathbf{v}_{1}$. 
Whilst in the kinematic analysis an arbitrary ideal flow may be superimposed on the particular solution, in this 
dynamic analysis  the zeroth-order (ideal) and first-order (non-ideal) Ohm's laws (\ref{eq:zerothohm}, \ref{eq:firstohm}) 
are coupled through the equation of motion (in particular equation~\ref{eq:secondmotion}). 
In the following section, Sec.~\ref{sec:composites}, we describe the circumstances in which an ideal flow
may be superimposed onto the same solutions for \eqref{eq:firstohm} that were detailed in paper I. We 
describe the nature  of these ideal flows and the consequences of their inclusion on the evolution of
magnetic flux.
Certain restrictions on the choice of ideal flow that may be superimposed on the particular solution
are found and so we in proceed, in Section~\ref{sec:accel}, %
to consider a more general ideal flow with a slightly different physical motivation.

\section{Composite Reconnection Solutions}
\label{sec:composites}

In general, the momentum equation given by \eqref{eq:secondmotion} implies a coupling  between the ideal 
and non-ideal Ohm's laws given by  equations \eqref{eq:zerothohm}  and \eqref{eq:firstohm}. However, for 
the class of ideal plasma flows $\mathbf{v}_{1}$  for which the curl of the inertial term on the left hand side of 
\eqref{eq:secondmotion}  vanishes, the equations become decoupled. In this case the effects of a non-trivial 
solution to \eqref{eq:zerothohm} are apparent at second order only in the form of the pressure gradient,  $\mathbf{\nabla}  p_{2}$. 
For ideal flows satisfying this condition the particular solutions of paper I may be taken as a solution to 
\eqref{eq:firstohm}, and so we have a direct comparison with the composite solutions of Hornig \& Priest (2003).

We begin in Section~\ref{subsec:simple} by examining an ideal stagnation flow $\mathbf{v}_{1}$  for which
\begin{math} \nabla \times \left( \mathbf{v}_{1} \cdot \nabla  \right) \mathbf{v}_{1} =0 \end{math} and 
continue in Section~\ref{subsec:flux} to consider the implications on the evolution of magnetic flux on both
this stagnation flow and other ideal flows with curl-free inertial terms.

\subsection{Simple Stagnation Flow}
\label{subsec:simple}

Clearly there are a wide variety of ideal plasma flows $\mathbf{v}_{1}$ for which 
\begin{math} 
\mathbf{\nabla}  \times \left( \mathbf{v}_{1} \cdot \mathbf{\nabla}   \right) \mathbf{v}_{1} = \mathbf{0} 
\end{math}
and are therefore candidates for the formation of composite solutions.
The different ideal flows each correspond to physically distinct scenarios. Stagnation flows 
are an obvious choice to consider since they enable the formation of  thin current sheets. In addition they 
allow for magnetic flux to be brought into and removed from the HFT and so, in contrast to the particular
reconnection solutions,  for the effects of the reconnection process to be seen on a wider scale.

\begin{figure}
\begin{center}
\subfigure[]{\label{fig:edge-a}\includegraphics[width=0.35\textwidth]{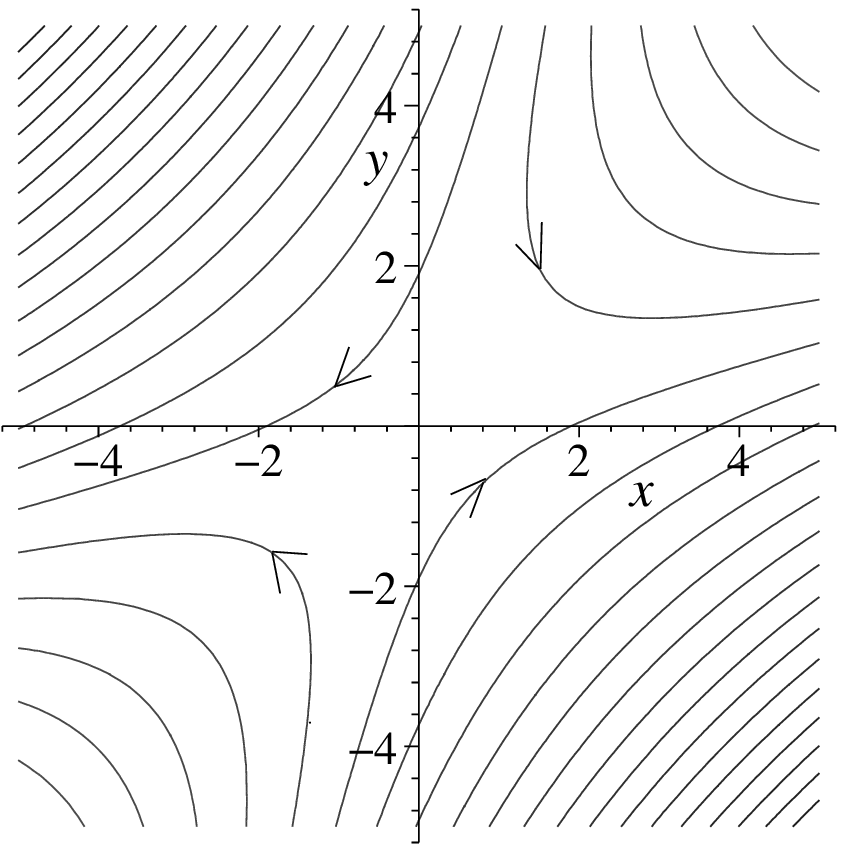}}
\subfigure[]{\label{fig:edge-b}\includegraphics[width=0.35\textwidth]{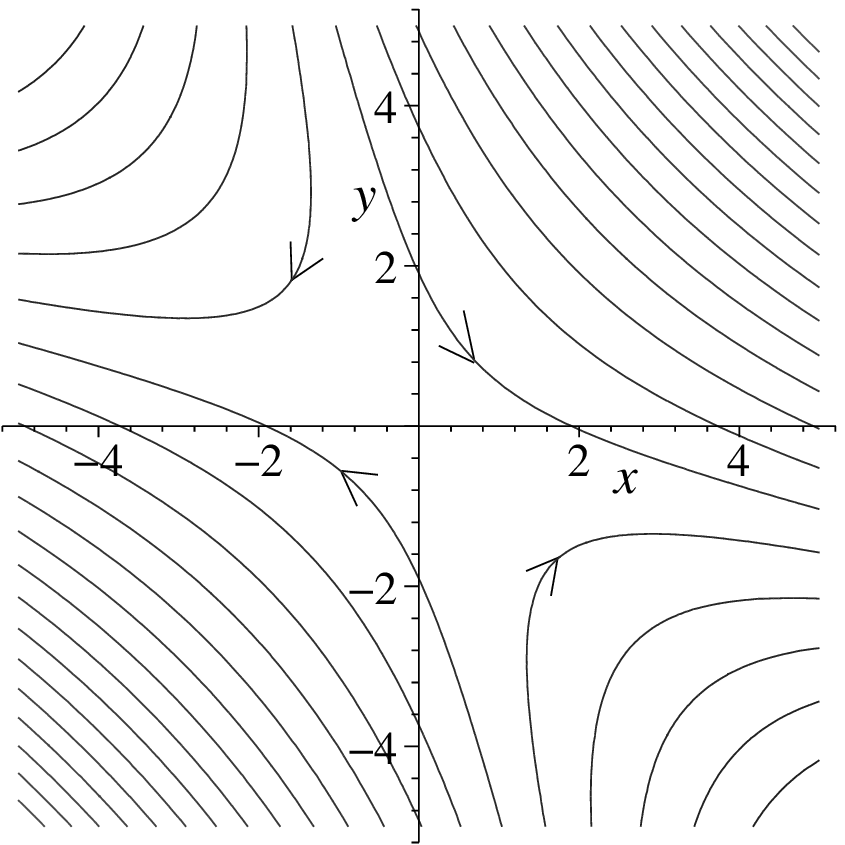}}
\end{center}
\caption{The ideal plasma velocity $\mathbf{v}_{1}$ for
(a) $z=0.5$ and (b) $z=-0.5$, and the parameters  $\varphi_{0}=1$, $k=0.5$, $b_{0}=2$, and $\Lambda=1$}
\label{fig:v1flows}
\end{figure}

We now turn to equation~\ref{eq:zerothohm} and determine a suitable ideal solution. We let $\phi_{0}$ be the
function of the field-line coordinates given by 
\begin{math} \left(x_{0}, y_{0}\right)\end{math} given by
\begin{equation}
\label{eq:simplephi0}
\phi_{0}=\frac{\varphi_{0}}{\Lambda^{2}} x_{0} y_{0}.
\end{equation}
Then, setting $Z_{0}=0$ the inverse field line mappings \eqref{eq:invfield} can be used
 to find an equivalent expression in terms of $x$, $y$, and $z$, so determining $\phi_{0}(x,y,z)$.
The component of \begin{math} \mathbf{v}_{1} \end{math} perpendicular to $\mathbf{B}_{0}$ 
may then be deduced from \eqref{eq:zerothohm}  as
\begin{equation} 
\label{eq:v1basic}
{\mathbf{v}_{1}}_{\perp}= - \frac{ \nabla \phi_{0} \times {\mathbf{B}_{0}} } {\vert {\mathbf{B}_{0}} \vert^{2}}, 
\end{equation} 
and the freedom in choosing a component parallel to  $\mathbf{B}_{0}$ exploited to set the 
$z$-component of  \begin{math} \mathbf{v}_{1} \end{math} to zero and so ensure the flow is divergence-free:
\begin{equation} 
\label{eq:howv1}
\mathbf{v}_{1}={\mathbf{v}_{1}}_{\perp}
           -\frac{ ( {\mathbf{v}_{1}}_{\perp} )_{z} \mathbf{B}_{0}}{b_{0}}.
\end{equation} 
Thus we obtain an ideal stagnation-type flow given by
\begin{displaymath}
\label{eq:v1}
\mathbf{v}_{1}=\frac{\varphi_{0}}{b_{0}\Lambda^{2}}
\left(x \cosh (2kz) - y \sinh (2kz) \right)  \boldsymbol{\hat{x}}
+\frac{\varphi_{0}}{b_{0}\Lambda^{2}}
\left(x \sinh (2kz) - y \cosh (2kz) \right)  \boldsymbol{\hat{y}},
\end{displaymath}
for which
\begin{math} \nabla  \times \left( \mathbf{v}_{1} \cdot \nabla  \right) \mathbf{v}_{1} =0 \end{math}, as required.
Some streamlines of the flow $\mathbf{v}_{1}$ are illustrated for a particular choice of 
parameter values in Figure~\ref{fig:v1flows}.

\begin{figure}
\begin{center}
\includegraphics[width=0.6\textwidth]{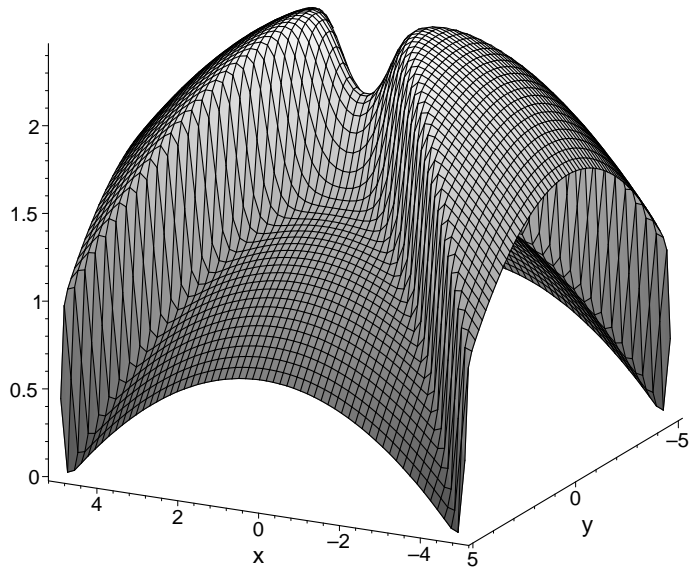}
\end{center}
\caption{The pressure profile $p_{2}\left(x,y\right)$ when $\mathbf{v}_{1}\neq\mathbf{0}$,
 $\mathbf{j}_{2}=j_{20} / \cosh^{2}\left(\left(x^{2}-y^{2}\right)/\lambda^{2}\right) \boldsymbol{\hat{z}},$ 
 and the parameters  $p_{20}=2$, $b_{0}=2$, $\lambda=1$, $k=1/2$, $j_{20}=-1$, $\Lambda=1$ and $\varphi_{0}=1/2$.
The lower pressure regions correspond to inflow of $\mathbf{v}_{1}$ and the higher pressure regions to outflow.}
 \label{fig:p2withv1}
\end{figure}

Since the inertial term $\left(\mathbf{v}_{1} \cdot \nabla  \right) \mathbf{v}_{1}$ may be expressed
as the gradient of a scalar function,
the equation of motion \ref{eq:secondmotion} has the same structure as that in the case of the particular
solutions examined in Paper I (where $\mathbf{v}_{1} = \mathbf{0}$). We therefore take the expressions
for $\mathbf{j}_{2}, \hat{\eta}, \mathbf{E}_{1}$ and $\mathbf{v}_{2}$ that were given in Paper I, Section 3.2.
The effect introducing the ideal-flow is seen in the pressure term $p_{2}$ which now becomes 
\begin{equation}
\label{eq:p2withv1}
p_{2}=p_{20} - 
 \mbox{\small{$\frac{1}{2}$}}
 k \lambda^{2} b_{0} j_{20} \tanh\left(\frac{x^{2}-y^{2}}{\lambda^{2}}\right)-
 \frac{ \varphi_{0}^{2}}{2 \Lambda^{2} b_{0}^{2} } 
 \frac{ \left(x^{2}+y^{2}\right)}{\Lambda^{2}}.
\end{equation}
The corresponding term for the particular solution may be recovered by setting $\varphi_{0}=0$, and
so it is seen that the inclusion of a zeroth-order flow has had the effect of introducing an extra 
term to the pressure, proportional to $\varphi_{0}^{2}/(\Lambda^{4}b_{0}^{2})$.  
When $\varphi_{0}=0$ there are strong gradients in the pressure  along the 
separatrices of $\mathbf{B}_{0}$ in the $xy$ plane. This extra term has the effect of 
smoothing out these strong gradients, with $p_{2}$ becoming a smoother function 
as $\varphi_{0}^{2}/\Lambda^{2}b_{0}^{2}$ is increased. An example of the pressure 
profile is shown in Figure~\ref{fig:p2withv1}, which can be compared with Figure~6 of Paper I.

The additional term has a natural physical explanation. It deflects the incoming 
$\mathbf{v}_{1}$ flow toward the outflow direction, a purely hydrodynamical effect.
Due to the symmetry with respect to inflow and outflow, there is no net transfer of
magnetic energy to kinetic (bulk) energy of the plasma in this stationary solution, as
 would be expected in a more realistic situation. However, we may model part of this 
process by requiring  $\mathbf{v}_{1} \cdot \mathbf{j}_{2} \times \mathbf{B}_{0}$
to be positive. This would result in an initial transfer of magnetic energy to 
kinetic energy, but with the latter subsequently transferred to potential energy,
since $\mathbf{v} \cdot \nabla p > 0$, so no net acceleration can take place.
We have here that
\begin{equation}
\label{eq:require}
\mathbf{v}_{1} \cdot \mathbf{j}_{2} \times \mathbf{B}_{0}=
-\frac{\varphi_{0} j_{20} k
\left( \left(x^{2}+y^{2}\right)\cosh\left(2kz\right) 
-2xy\sinh\left(2kz\right)\right)
} { \Lambda^{2}   \cosh^{2}\left(\frac{x^{2}-y^{2}} {\lambda^{2}} \right)}.
\end{equation}
In order to model the situation in which the Lorentz forces drive the reconnecting plasma
flow we require the expression \eqref{eq:require} to be positive, which can be ensured by taking
the combination $\varphi_{0}j_{20}k<0$.  Note, however, that cases where external pressure differences
drive flows against Lorentz forces (`driven-reconnection') are also known.

The reconnection rate in the model is given by the integral of the parallel 
electric field along the reconnection line which is identified with the $z$-axis:
\begin{eqnarray}
\label{eq:recorate}
\frac{ d \Phi_{\textrm{mag}} } {dt} & = & \int \mathbf{E_{\parallel}} \ dl
= \int \frac{\left(\mathbf{E}_{0} \cdot \mathbf{B}_{0} + 
M_{e} \mathbf{E}_{1} \cdot \mathbf{B}_{0} + O(M_{e}^{2})\right)} 
{\vert \mathbf{B}_{0} \vert} dl  \notag \\
 & = & 
 \sqrt{\pi} M_{e} \ e_{10} L +  O\left(M_{e}^2\right).
\end{eqnarray}
Since we have taken the same non-ideal electric field $\mathbf{E}_{1}$ in both the 
particular solutions of Paper I and these composite solutions the expression for the 
rate of reconnection is clearly the same in both cases. However, the non-vanishing
external ideal flow changes the interpretation of this reconnection rate since the
reconnection process can now reconnect flux lying initially outside  of the 
HFT,  as will be demonstrated in the following section.

We could, in principle, determine higher-order quantities of the expansion solution using
the iterative scheme outlined in Paper I. In the particular solution we found that Ohm's law
at subsequent even-orders and the equation of motion at subsequent odd-orders could be
set to zero but in this case the remaining equations must all be solved numerically.

\subsection{Magnetic Flux Evolution}
\label{subsec:flux}

We have shown that the introduction of the ideal stagnation flow 
\begin{math} \mathbf{v}_{1} \end{math}, as described in the previous section, does not alter the 
reconnection rate (up to second order). The evolution of magnetic flux in the two cases is, however, 
quite different, and may be visualised using the concept of a  magnetic flux velocity (e.g. Hornig 
\& Schindler, 1996).  Such a velocity satisfies
\begin{displaymath}
\mathbf{E} + \mathbf{w} \times \mathbf{B} = \mathbf{0}.
\end{displaymath}
In two-dimensions, a non-vanishing electric field at an X-point of the magnetic field requires a singularity in 
the flux velocity at that point.  The singularity describes the reconnection taking place at that point. In 3D, on 
the other hand, there is no general well-defined flux-transport velocity. However, provided the non-ideal region 
is localised and contains no closed flux, we can replace the notion of a single flux-transport velocity by a pair
of velocities $\mathbf{w}_{in}$ and $\mathbf{w}_{out}$ that describe the behaviour of magnetic flux entering
and leaving the non-ideal region, respectively (Priest, Hornig \& Pontin, 2003). The two flux velocities will not 
coincide within the non-ideal  region and through examining the difference between the two we may deduce 
information about the behaviour of magnetic flux in the process.  

\begin{figure}
\begin{center}
\includegraphics[width=0.5\textwidth]{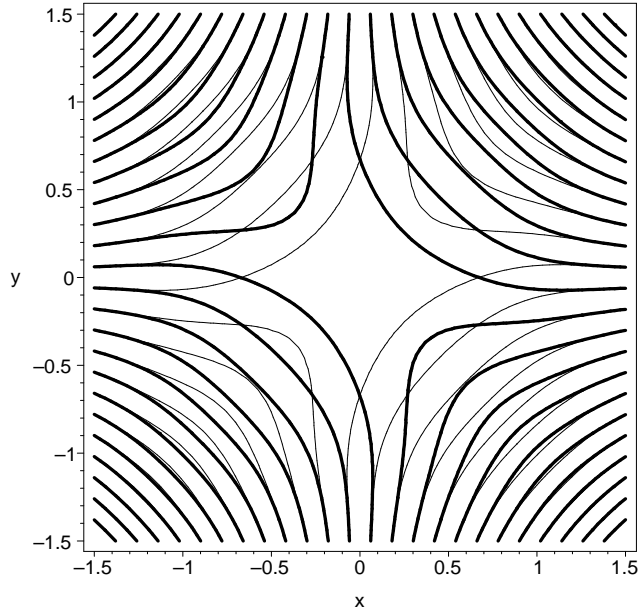}
\end{center}
\caption{Flow $\mathbf{w}_{in}$ (grey) and $\mathbf{w}_{out}$ (black) for the 
solution described in Section~\ref{sec:composites}(a).}
\label{fig:contours1}
\end{figure}

\begin{figure}
\begin{center}
\includegraphics[width=0.4\textwidth]{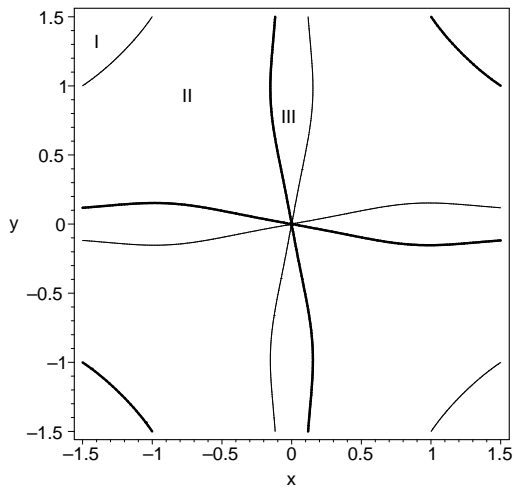}
\end{center}
\caption{Separatrices of  $\mathbf{w}_{in}$ (grey) and $\mathbf{w}_{out}$ black for the solution described in 
Section~\ref{sec:composites}(a). The region is divided into three types of reconnective behaviour.
Magnetic flux passing through region I undergoes ideal evolution.
Magnetic flux passing through region II undergoes a slippage-like behaviour while flux passing
through region II undergoes an evolution similar to that seen in classical 2D reconnection}
\label{fig:separ1}
\end{figure}

The use of these flux velocities therefore allows us to track the evolution of plasma elements in the ideal region above
and below the localised non-ideal region.  Initially-connected elements will only remain connected 
if the field line linking them does not pass through the non-ideal region; otherwise the elements will change their 
connectivity. The  flow-lines corresponding to the ideal evolution above and below the non-ideal
region can be projected onto the central plane, say, using the pair of quasi-velocities 
\begin{math} \mathbf{w}_{in} \end{math} and \begin{math} \mathbf{w}_{out} \end{math}. For the stagnation 
flow described in the previous section, the relevant projection is shown in Figure~\ref{fig:contours1}
(for a particular choice of parameter values).  The flow lines of 
\begin{math} \mathbf{w}_{in} \end{math} (grey lines) in the $z=0$ plane are 
superimposed on those of \begin{math} \mathbf{w}_{out} \end{math}  (black lines) 
in the same plane. We are able to divide the plane into three regions according
to the type of reconnective behaviour that occurs; the separatrices dividing these
regions are shown in Figure~\ref{fig:separ1}.

In region I the flow lines of \begin{math} \mathbf{w}_{in} \end{math} and 
\begin{math} \mathbf{w}_{out} \end{math} coincide perfectly. The magnetic 
flux passing through the $z=0$ plane in region I evolves ideally, so that
initially-connected plasma elements will remain connected. In regions II 
and III the flow lines of \begin{math} \mathbf{w}_{in} \end{math} and 
\begin{math} \mathbf{w}_{out} \end{math} do not coincide. 
For magnetic flux passing through the $z=0$ plane in these two regions
we deduce that plasma elements above and below the non-ideal region that 
are initially connected will not remain so. Tracking the evolution of 
corresponding pairs allows us to distinguish different types of magnetic 
flux evolution. 

Magnetic flux passing through region II exhibits a slippage-like behaviour.
Initially connected plasma elements above and below the non-ideal region
will change their connections as the flow transports the magnetic flux 
linking them into the non-ideal region.  On leaving the shadow of the non-ideal region
the initially connected elements are both transported in the same direction by the flow 
and a new ideal connection is again established for each plasma element. Although
this connnection will not be with the initial partner, it will be
with a plasma element that was initially close to that partner. 

Magnetic flux passing through region III exhibits the type of behaviour
most similar to that shown in classical 2D reconnection. Again, initially-connected 
plasma elements above and below the non-ideal region loose their
connections as the magnetic flux linking them is transported into the
non-ideal region. However, on leaving the shadow of the non-ideal region
the initially-connected plasma elements above and below the non-ideal region
are transported in different directions by the flow, along  opposing
`wings' seen in Figure~\ref{fig:separ1}, and their separation will
therefore increase in time, as in the classical 2D reconnection picture. 
The new ideal connection for a plasma element initially above (below) the 
non-ideal region will be with a plasma element that was initially below (above) the 
non-ideal region in the distant opposing wing.

\begin{figure}
\begin{center}
\includegraphics[width=0.4\textwidth]{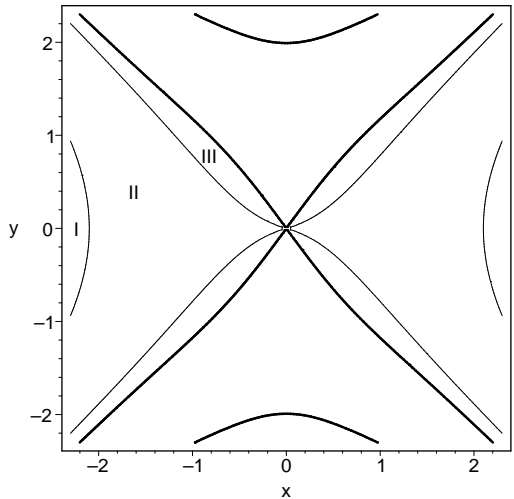}
\end{center}
\caption{Separatrices of  $\mathbf{w}_{in}$ (grey) and $\mathbf{w}_{out}$ black when $\mathbf{v}_{1}$
is given by \eqref{eq:not_crossing}. The same three types of reconnective behaviour as Figure~\ref{fig:separ1}
 are present}         
\label{fig:separ2}
\end{figure}

We have been able to make a direct comparison between the pure solutions described in Paper I and the
composite solutions described here since, for our particular choice of $\mathbf{v}_{1}$, the curl of
the inertia term in \eqref{eq:secondmotion} vanishes. 
In principal we could have chosen other ideal flows for this directly comparable analysis that
also have a curl-free inertial term. One example is that which results from defining the scalar function
$\phi_{0}$ as the function of field-line coordinates given by
\begin{displaymath}
\phi_{0}=\frac{\varphi_{0}}{\Lambda^{2}}\left(x_{0}^{2}- y_{0}^{2}\right),
\end{displaymath}
from which we obtain
\begin{equation}
\label{eq:not_crossing}
\mathbf{v_{1}} = \frac{-2\varphi_{0}}{b_{0}\Lambda^{2}} \left(y \boldsymbol{\hat{x}} + x\boldsymbol{\hat{z}}\right).
\end{equation}
This is also a stagnation flow, but it differs considerably from the flow considered 
in the previous section; it does not cross the separatrices of the projection of 
$\mathbf{B_{0}}$ onto the $xy$-plane,  and is independent of the third coordinate, $z$.
When superimposed on the particular solution, however, the same three regions of differing
flux evolution are present, as illustrated in Figure~\ref{fig:separ2}. 
The inflow and outflow channels bounded by the separatrices of the quasi-flux velocities 
are now centred around the separatrices of 
 $\mathbf{B}_{0}$ in the $z=0$ plane. This demonstrates one of the crucial differences
between 2D and 2.5D reconnection and the 3D case. The crossing of the separatrices 
by the flow is only a criterion for reconnection in the 2D case. In 3D the difference
between $\mathbf{w}_{in}$ and $\mathbf{w}_{out}$ is the crucial property for reconnection.

Another example in this class of flows which can be used to form composite solutions is the rotational 
ideal flow arising from the choice
\begin{equation}
\label{eq:rotate}
\phi_{0}=\frac{\varphi_{0}}{\Lambda^{2}}\left(x_{0}^{2}+ y_{0}^{2}\right).
\end{equation}
This ideal flow is rotating in the same sense for all $z$, and so does not have 
the effect of bringing flux into and away from the non-ideal region.

For the three flows examined in this section, the reconnection rate, as determined by the
integral of the parallel electric field along the reconnection line, is identical,
but the magnetic flux evolution quite different. 
The distinct types of reconnective behaviour illustrated here, and in
Paper I, may be distinguished by considering the associated internal and external reconnection
rates, as introduced by Hornig (2007).
\begin{equation*}
\frac{d\Phi_{\rm rec}}{dt}= \frac{d\Phi_{\rm rec}^{\rm external}}{dt} + \frac{d\Phi_{\rm rec}^{\rm internal}}{dt} .
\end{equation*}
The external reconnection rate measures the rate at which ßux is transported into (and, equivalently,
  out of) the non-ideal region. This rate is always less than or equal to the total reconnection rate. The difference between the two is the internal reconnection rate which is non-zero only if  there is  a circular flow around the reconnection line rather than a stagnation flow.


The stagnation flow examples illustrated in Figures~\ref{fig:separ1} and \ref{fig:separ2} both
correspond to a purely external reconnection rate. In this situation the separatrices of 
the flow (which divide regions II and III) pass though the origin, and so the difference 
in the electric potential between them is equal to the total reconnection rate, 
i.e. the difference in electric potential across the non-ideal region.
For an ideal rotational flow, such as that arising from the electric potential
given by \eqref{eq:rotate}, the external reconnection rate vanishes and the 
reconnection is internal only. Similarly if the ideal flow is zero (as in the case of
the particular solutions of Paper I) then the reconnection is purely internal.
Thus the interpretation of reconnection rate in this way allows for a clear 
distinction between the different types of solutions considered here.

We note also that a combination of internal and external reconnection is not excluded
in these solutions, and is expected if a smooth transition between the purely external reconnection 
solutions illustrated in figures~\ref{fig:separ1} and \ref{fig:separ2} and the purely
internal reconnection found in the particular solution is to be made.
 Such a solution exists when the magnitude
of the ideal flow \begin{math} \mathbf{v}_{1} \end{math} is decreased to be the same,
or less than, that of the non-ideal flow \begin{math} M_{e} \mathbf{v}_{2} \end{math}. 
In addition to the three regions of differing space flux evolution described above and illustrated
in Figure~\ref{fig:separ1}, the magnetic flux in these mixed solutions would show rotational 
dynamics within part of the HFT.

\section{Accelerating Stagnation Flow}
\label{sec:accel}

In a realistic situation we would expect to see a plasma flow that results in a 
net transfer of magnetic energy to kinetic (bulk) energy of the plasma since magnetic 
energy is the main source of energy in the solar corona.  
This property must be explicitly prescribed here since the model does not
include the time-dependent processes external (and possibly prior) to the reconnection
process that lead to the build-up of a current sheet and corresponding plasma flows.
Instead these properties are represented in the model via boundary conditions on the
flow, magnetic field and pressure profiles.
Thus in the expansion scheme we 
may ensure an increase in kinetic energy occurs in the reconnection process by requiring
\begin{math} 
\mathbf{v}_{1} \cdot \mathbf{j}_{2} \times \mathbf{B}_{0} -
\mathbf{v}_{1} \cdot \nabla' p_{2}
\end{math}
to be, on average, positive over the volume (which is not the case for the above stagnation flow).
This increase in kinetic energy may be the result of a transfer of magnetic energy
(due to \begin{math} \mathbf{v}_{1} \cdot \mathbf{j}_{2} \times \mathbf{B}_{0} \end{math}),
a transfer of thermal energy (due to \begin{math} \mathbf{v}_{1} \cdot \nabla' p_{2} \end{math}) or
a combination of both effects.  Numerical experiments  suggest 
 considering a plasma flow that sharply changes its direction toward the outflow region;
in such experiments fast jets of plasma emerging from the reconnection region are observed.
We examine in this section an ideal plasma flow, $\mathbf{v}_{1}$, which possesses
these properties, using a method similar to that of Section~(\ref{subsec:simple}). 
Now, however, since the curl of the inertial term in \eqref{eq:secondmotion} does 
not vanish, there is a much larger degree of coupling between equations 
\eqref{eq:zerothohm} and \eqref{eq:firstohm}, and the particular solution
of Paper I  
can no longer be used as a solution to \eqref{eq:firstohm}.

\begin{figure}
\begin{center}
\includegraphics[width=0.45\textwidth]{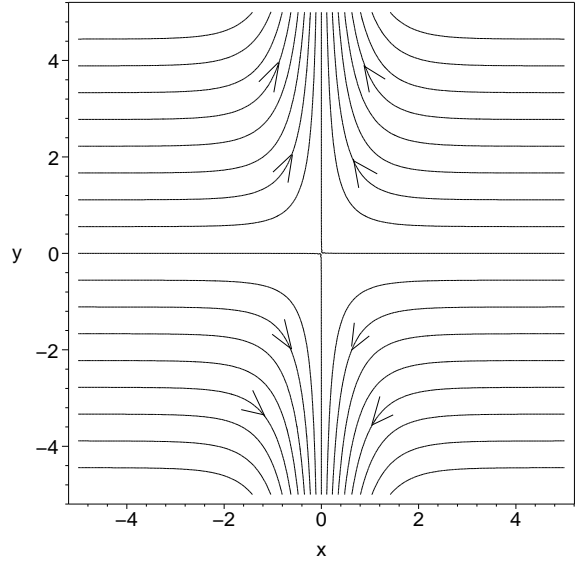}
\end{center}
\caption{Stagnation-point structure of the
 velocity field $\mathbf{v_{1}}$ corresponding to $\phi_{0}=-\varphi_{0} y_{0}\tanh(x_{0})/\Lambda^{2}$
in the plane $z=0$.  As indicated by the closeness of the contours, the plasma has a 
greater velocity along the outflow direction.}
\label{fig:v1singlejet}
\end{figure}

Just as in the previous section we will consider incompressible solutions; a 
plasma flow with a faster outflow velocity than inflow velocity must have an 
associated outflow channel that is narrower than its inflow channel. To achieve 
such a flow we impose a non-symmetric function,  $\phi_{0}$, for the lowest order 
electric potential, and then deduce the  plasma velocity $\mathbf{v}_{1}$  
from \eqref{eq:zerothohm}. For example, we may impose  $\phi_{0}$ as the 
function of field line coordinates  given by
\begin{equation}
\label{eq:singlejet}
\phi_{0}=-\frac{\varphi_{0}}{\Lambda^{2}} y_{0}\tanh\left(x_{0}\right),
\end{equation}
and use the inverse field line mappings to deduce an equivalent expression in terms of $x$, $y$ and $z$.
An analytical expression for  \begin{math} \mathbf{v_{1}} \end{math} (which is too long to be shown here)
is found using \eqref{eq:howv1}. In the central region the flow has a stagnation structure, as shown
 in Figure~\ref{fig:v1singlejet}, with single inflow and outflow channels that are of different widths. 
Thus, depending on the direction of the flow, and since it is incompressible, an acceleration or deceleration
 of the plasma takes place. The physically relevant case corresponds to the choice  $\varphi_{0}>0$, for which 
the outflow direction is the narrower channel along the $y$-direction, and so the plasma is accelerated during
 the reconnection process.


Turning now to the lowest-order momentum equation in the expansion scheme,  \eqref{eq:secondmotion},
we integrate along the  field lines 
\eqref{eq:fieldlines}, starting from the plane $z=0$ to deduce the pressure $p_{2}$:
\begin{equation}
\label{eq:intforp}
p_{2}\left(x,y,z\right) =
 - \int_{s=0}^{z/b_{0}} \left(\mathbf{v}_{1} \cdot \nabla' \   \mathbf{v}_{1} \right) \cdot \mathbf{B}_{0} \  ds
+p_{2}\left(x_{0},y_{0}\right).
\end{equation}
We first examine solutions obtained when the free function $p_{2}\left(x_{0},y_{0}\right)$ is set to zero.
 Later in the section we shall consider another particular example where $p_{2}\left(x_{0},y_{0}\right) \neq 0$,
and show that the choice of this free function has a considerable effect on the reconnection process.

\begin{figure}
\begin{center}
\subfigure[]{\label{fig:edge-f}\includegraphics[width=0.4\textwidth]{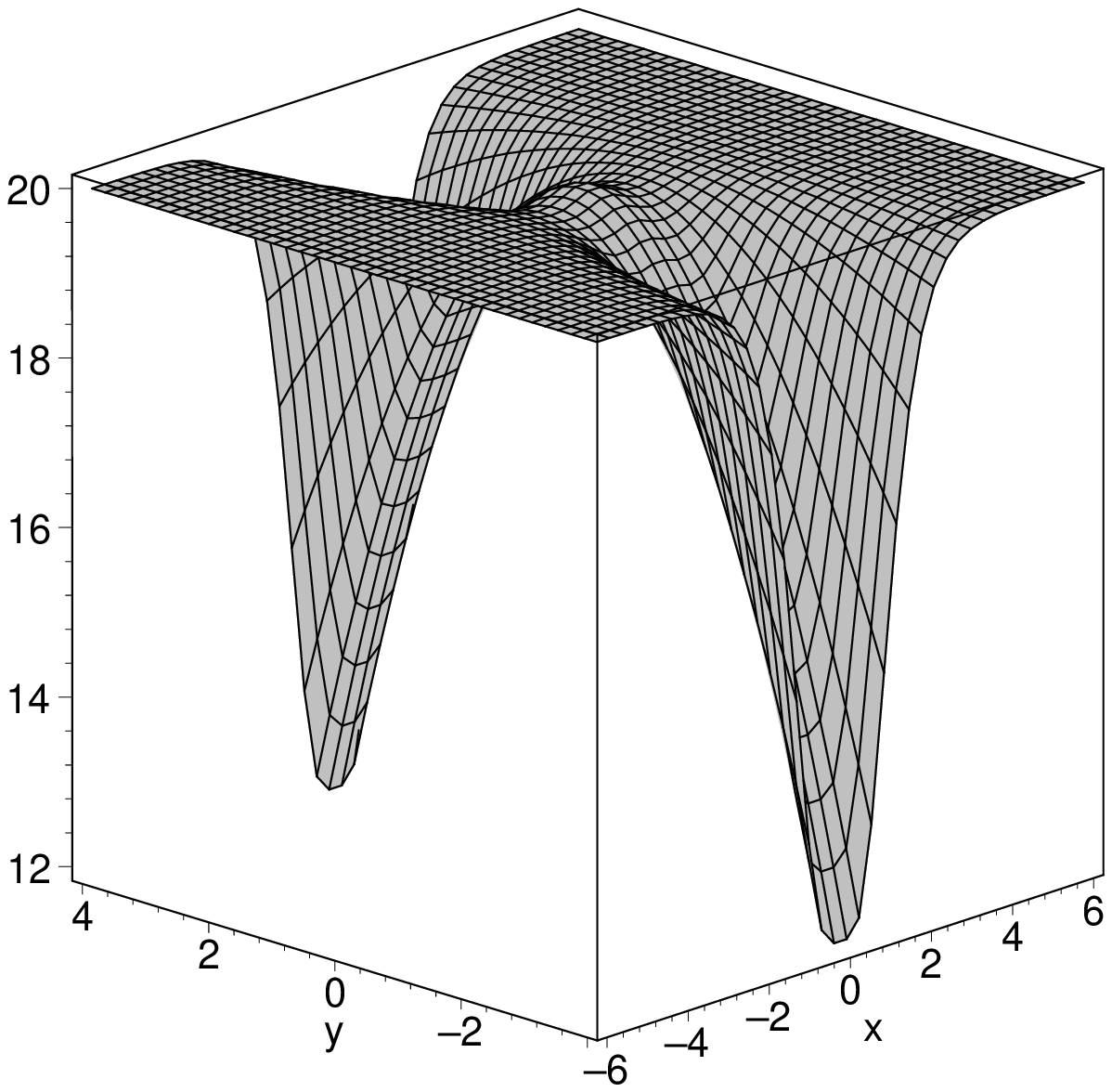}}
\subfigure[]{\label{fig:edge-g}\includegraphics[width=0.38\textwidth]{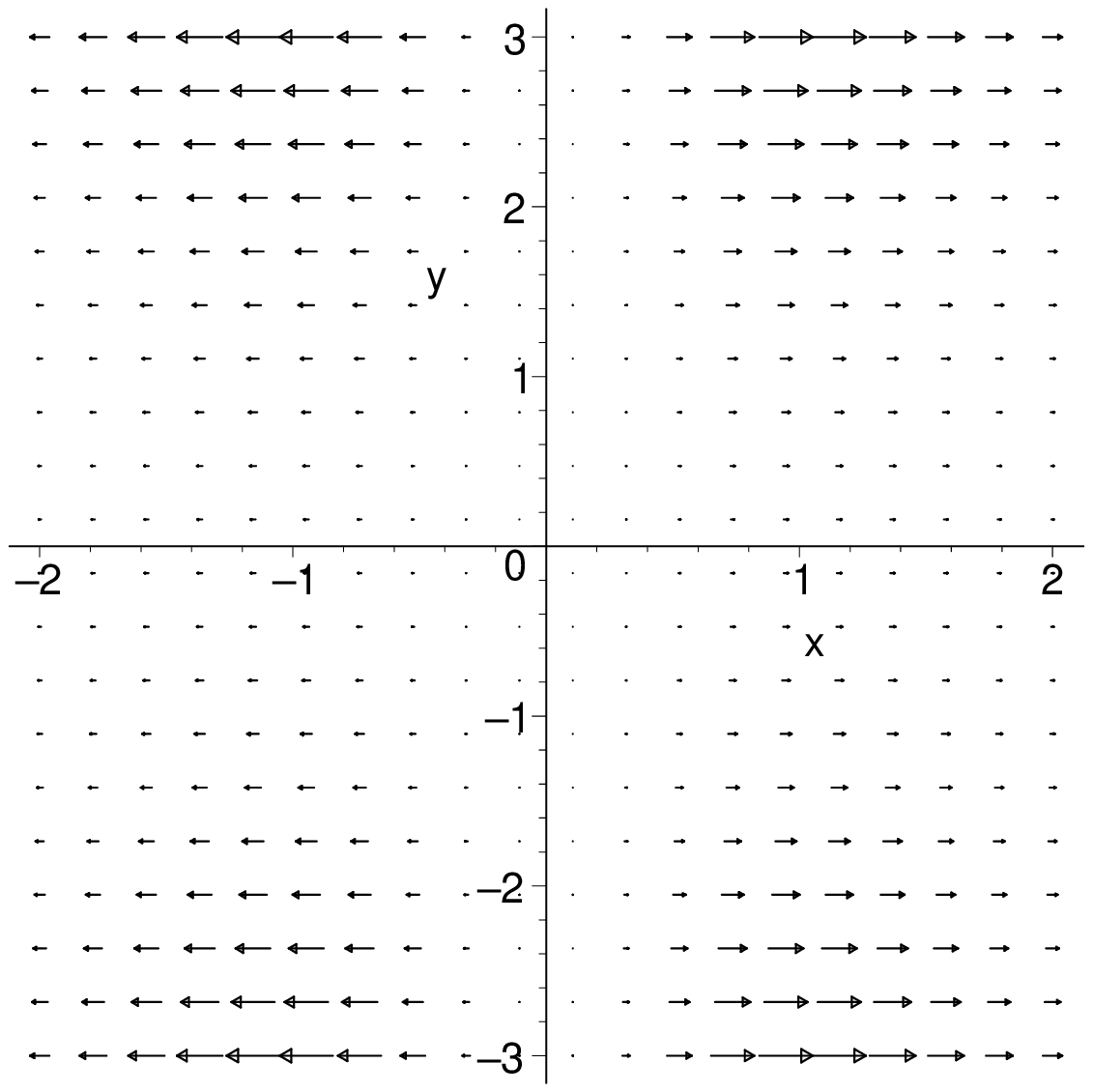}}
\end{center}
\caption{(a) The pressure profile $p_{2}$ and (b) the Lorentz force $\mathbf{j}_{2} \times \mathbf{B}_{0}$
in the plane $z=0$ for an accelerating stagnation
 flow $\mathbf{v}_{1}$. The free function $p_{2}\left(x_{0},y_{0}\right)$ in 
equation~\eqref{eq:intforp} has been set to zero.  The outflow is aligned with the
$y$-axis, and corresponds to the channel of decreasing pressure.}
\label{fig:singlep2}
\end{figure}

An example of the pressure profile in the case where \begin{math}p_{2}\left(x_{0},y_{0}\right)=0\end{math}
is shown in Figure~\ref{fig:singlep2}(a). The expression obtained for $p_{2}$ 
 is dependent on $\varphi_{0}^{2}$, and so the pressure profile is independent of the flow direction.
Thus for the case $\varphi_{0}>0$ which we are considering here, a pressure gradient exists along
the outflow direction which is in the direction of the flow, and so acts to accelerate the plasma.

The perpendicular component of the current, \begin{math} \mathbf{j_{2_{\perp}}} \end{math}, can be
 determined analytically from \eqref{eq:secondmotion} once the pressure is given:
\begin{displaymath}
{\mathbf{j}_{2}}_{\perp}=\frac{
\left(-\nabla' p_{2} -  \left(\mathbf{v}_{1} \cdot \nabla' \right) \mathbf{v}_{1} \right) \times \mathbf{B}_{0}}
{ \vert \mathbf{B}_{0} \vert^{2}}.
\end{displaymath}
A Lorentz force is present within the outflow channels, and is 
directed away from the central line of minimum pressure, as shown in 
Figure ~\ref{fig:singlep2}(b).  Thus, since it is not aligned with the flow 
direction, this force does not act to alter the plasma velocity; in this
example it is only the pressure gradient which  accelerates the plasma, 
causing the fast outflow jets. The quantity
\begin{math} 
\mathbf{v}_{1} \cdot {\mathbf{j}_{2}}_{\perp} \times \mathbf{B}_{0} -
\mathbf{v}_{1} \cdot \nabla' p_{2}
\end{math}
is, on average, positive over the region provided $\varphi_{0}>0$, i.e. provided the flow is accelerated from 
its inflow to outflow direction. This net acceleration, a consequence of the pressure gradient,
 results in a net transfer of thermal energy to kinetic energy.


The full form of the current \begin{math} \mathbf{j}_{2} \end{math} may be determined by
finding a scalar function \begin{math} \lambda\left(x,y,z\right) \end{math}
 such that setting \begin{math} \mathbf{j}_{2}= {\mathbf{j}_{2}}_{\perp} + \lambda \mathbf{B}_{0} \end{math}
ensures the current is divergence-free. Taking
\begin{math} \nabla' \cdot {\mathbf{j}_{2}}_{\perp} + \nabla' \lambda \cdot \mathbf{B}_{0} =0 \end{math} and
integrating along the field lines gives
\begin{eqnarray}
\label{eq:j2int}
\lambda(x,y,z) & = & -\int_{s=0}^{s=z/b_{0}} \nabla'  \cdot\mathbf{j_{2}}_{\perp} \ ds 
                                                        + \lambda\left(x_{0},y_{0}\right) \\
               & = & \tilde{\lambda}(x,y,z) + \lambda(x_{0},y_{0}), \notag
\end{eqnarray}
where $\lambda\left(x_{0},y_{0}\right)$ is a function that we are free to impose on the solution
The current is then given by
\begin{displaymath} 
\mathbf{j}_{2} = ( \ {\mathbf{j}_{2}}_{\perp} + \tilde{\lambda}(x,y,z) \mathbf{B}_{0} )
                       + \lambda(x_{0},y_{0}) \mathbf{B}_{0}
               = \tilde{\mathbf{j}}_{2} + \mathbf{j}_{2}^{\ast}, 
\end{displaymath}
where $\mathbf{j}_{2}^{\ast}$ is solely determined by the free function $\lambda\left(x_{0},y_{0}\right)$.
The term $\mathbf{j}_{2}^{\ast}$ also determines the current along the $z$-axis because,
 due to the vanishing divergence of ${\mathbf{j}_{2}}_{\perp}$ along the $z$-axis, 
the $z$-component of $\tilde{\mathbf{j}}_{2}$ vanishes there.
 Equation~\eqref{eq:firstohm} then implies that the reconnection rate will be 
determined by this free function (together with the form of $\hat{\eta}$), 
rather than by the ideal flow \begin{math} \mathbf{v}_{1} \end{math}.
 i.e.~governed by the lowest order non-ideal solution.


%


%

As previously mentionned,
we can make use of the freedom to choose the function  
\begin{math} p_{2}\left(x_{0},y_{0}\right) \end{math} 
that arises in the integration for $p_{2}$, given by equation~\eqref{eq:intforp}.
The form taken for \begin{math} p_{2}\left(x_{0},y_{0}\right) \end{math} 
will alter the pressure profile, and consequently 
also the current \begin{math} \mathbf{j}_{2} \end{math}.
Thus the various choices correspond to
additional 3D reconnection solutions with differing physical
motivations.
In particular, a form for \begin{math} p_{2}\left(x_{0},y_{0}\right) \end{math} 
may be imposed such that
the acceleration of the plasma is driven by the Lorentz force, 
\begin{math} \mathbf{j}_{2} \times \mathbf{B}_{0} \end{math},
rather than by the pressure gradient, \begin{math} -\nabla' p_{2} \end{math}.
\begin{figure}
\begin{center}
\subfigure[]{\label{fig:edge-h}\includegraphics[width=0.38\textwidth]{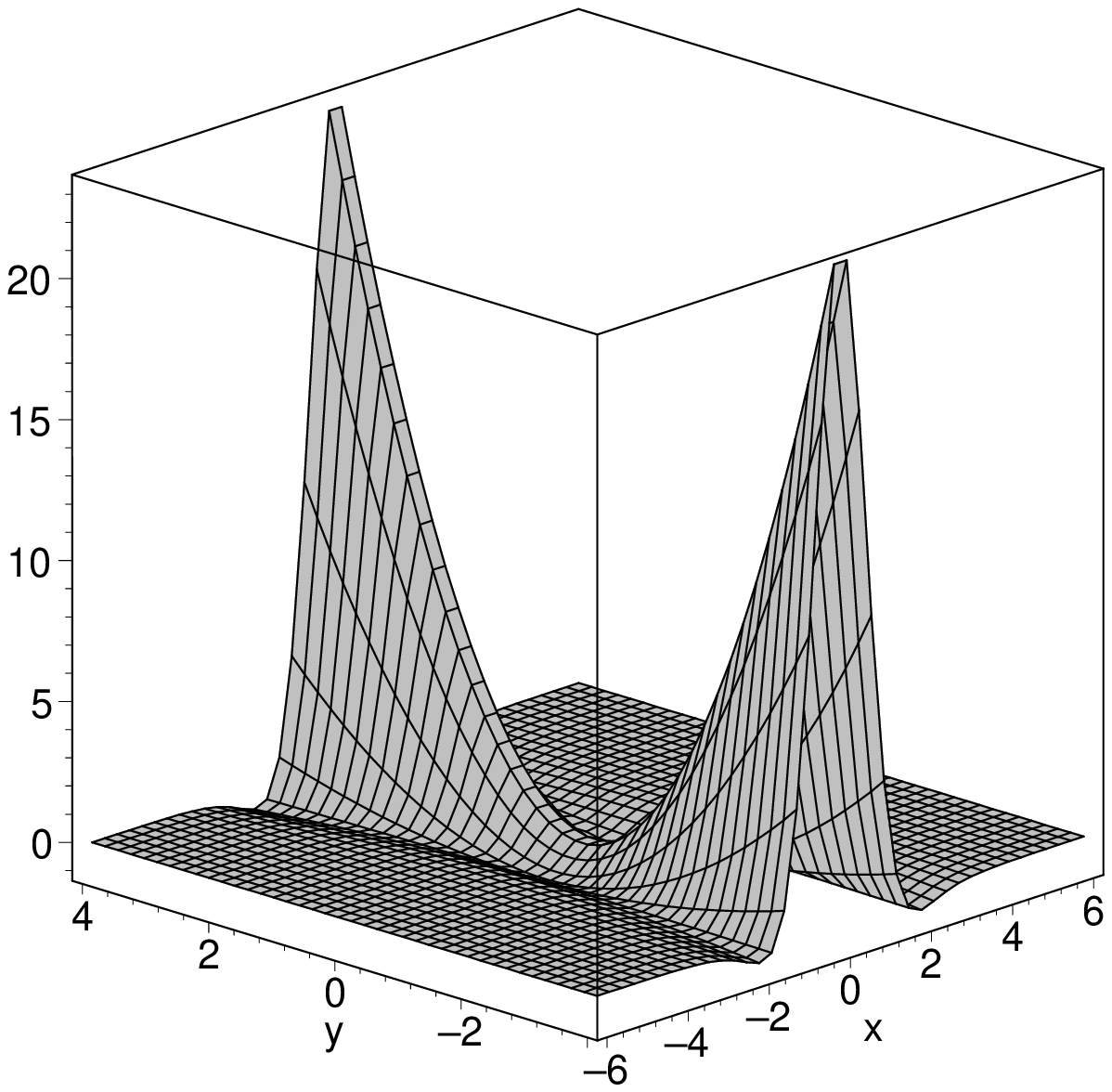}}
\subfigure[]{\label{fig:edge-i}\includegraphics[width=0.38\textwidth]{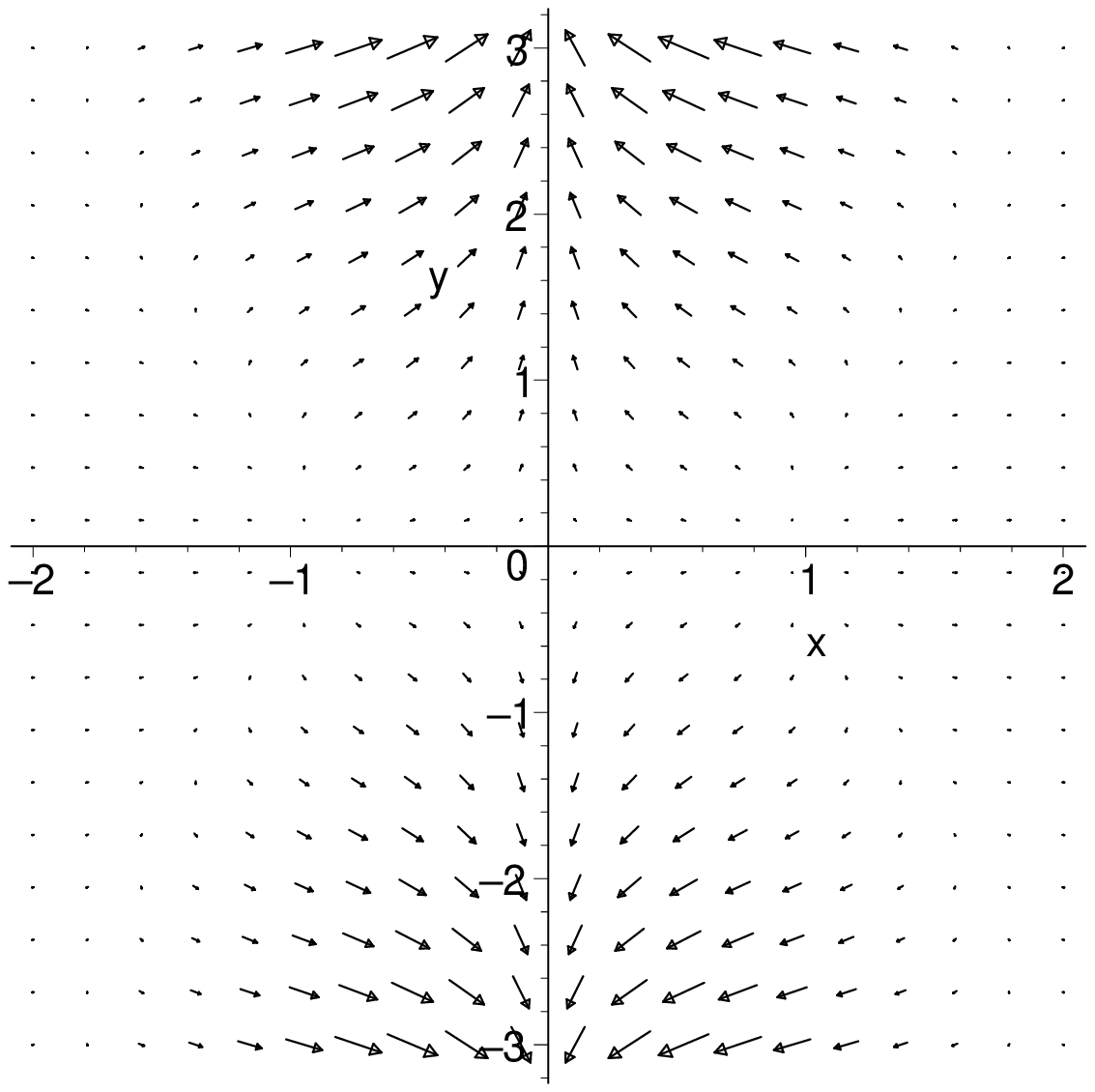}}
\end{center}
\caption{(a) The pressure profile $p_{2}$ and (b) the Lorentz force $\mathbf{j}_{2} \times \mathbf{B}_{0}$
for the accelerating stagnation flow, now with free function $p_{2}\left(x_{0},y_{0}\right)$ 
in equation~\eqref{eq:intforp} given by \eqref{eq:freebit}.
 The outflow is aligned with the $y$-axis, and corresponds to the channel of increasing pressure.
This figure may be compared with Figure~\ref{fig:singlep2}.}
\label{fig:NOp20singlep2}
\end{figure}
One such example is obtained by adding the additional function given (in terms of the
fieldline coordinates) by
\begin{equation}
\label{eq:freebit} 
p_{2}\left(x_{0},y_{0}\right) = p_{20} \  e^{-x_{0}^{2}} \left(y_{0}^{2} -x_{0}^{2}\right) 
\end{equation}
to the pressure $p_{2}$. The resultant pressure profile is illustrated in Figure~\ref{fig:NOp20singlep2}(a), 
where it is seen that the pressure gradient is still directed along the outflow channel, but now acts against the
direction of the flow. Therefore, the acceleration of the plasma is driven entirely by the Lorentz force,
with magnetic energy being transfered to kinetic energy in the reconnection process.
The Lorentz force is illustrated in Figure~\ref{fig:NOp20singlep2}(b), where the 
perpendicular component of the current \begin{math} \mathbf{j}_{2} \end{math} has been deduced using the
same method as described above.
The divergence of ${\mathbf{j}_{2}}_{\perp}$ along the $z$-axis remains zero 
with the inclusion of the extra factor in the solution for $p_{2}$. Thus the reconnection
rate in this case is still determined completely by the free function  $\lambda\left(x_{0},y_{0}\right)$,
and may therefore be the same as in the previous pressure gradient driven model.

In both of the examples in this section we must impose a localised resistivity in order to ensure
a localised non-ideal region. In principle the remaining quantities could then determined numerically using
the iterative scheme outlined in Paper I.

\section{Conclusions}
\label{sec:concs}

In this series of papers we have focussed on the fundamental questions of existence and
parametric dependence of steady-state 3D MHD magnetic reconnection.
Through the use of an expansion scheme, a number of examples have been provided that illustrate 
the very wide range of solutions possible in 3D, a key feature of 3D reconnection.  At each level of the expansion 
several free parameters  and functions appear, with the various possible choices for each of these corresponding 
to physically different 3D reconnection events.


One aspect of the diversity in 3D solutions was suggested by the kinematic analysis of Hornig \& Priest (2003)
where it was noted that in 3D, for a given magnetic field, Ohm's law can be decomposed into non-ideal
and ideal components, the non-ideal solution being termed `particular' and the sum of the two `composite'. 
The analysis presented here demonstrates a similar decomposition can be made in a fully dynamic situation
with the non-ideal and ideal solutions being linked together through the equation of motion.

In Paper I particular solutions were analysed in a setup with the non-ideal region localised in all 3D 
and the basic state an X-type current-free equilibrium. The plasma flow in these solutions is purely rotational
and confined to be within the HFT consisting of all the field-lines threading the non-ideal region. 
In Section~\ref{sec:composites} of the present work we considered the extent to which the coupling between the non-ideal
and ideal flows through the equation of motion restricts the form of the ideal solution and examined the 
effect of the ideal solution on the reconnection rate, evolution of flux and energetics.
A particular class of ideal flows (for which the inertial term  can be expressed as a 
gradient), may be imposed on the particular non-ideal solution  without altering the form of the current,  parallel 
electric field or (in consequence) the reconnection rate. For these flows the coupling between the two 
solutions is relatively  weak, affecting only the pressure term in the non-ideal solution.  A wide range of flows, 
both in strength and, more importantly, in profile, belong to this class of solution. They may be distinguished 
by their effect on the  evolution of magnetic flux.

In general, stagnation flows are expected to be present if classical reconnection is to occur, since they allow 
thin current sheets to be built up  and so localised non-ideal regions to become established. A variety of 
symmetric  stagnation flows, as considered in Section~\ref{sec:composites}, belong to the class of ideal flows 
that may be used to form composite solutions. These flows bring magnetic flux into the non-ideal region from 
large distances and subsequently remove the flux. Magnetic flux threading particular channels in the centre of 
the region shows similar behaviour to typical 2D reconnection, in the sense that field lines
brought in toward the non-ideal region reconnect with field lines initially far away, and the separation 
of initially connected plasma elements increases in time after the flux has left the non-ideal region.
In the same reconnection event, magnetic flux  passing through other regions of the domain can be seen to 
undergo a slippage-like behaviour. Although the reconnection rate in the particular and composite reconnection
solutions is quantitatively the same, its physical interpretation differs. 
The reconnection, which was completely 
internal for the particular solution is now completely external for these stagnation flows. 
It was also shown that the 2D criterion for reconnection, namely that the plasma flow has to 
cross the separatrices, does not hold in 3D since we can construct solutions which do 
not have this property but nevertheless have a non-vanishing reconnection rate.


In these symmetric examples there is no net transfer of magnetic energy to bulk energy of the plasma. 
Non-symmetric stagnation flows, however, such as those  considered in  Section~\ref{sec:accel} can 
convert magnetic energy  into kinetic energy. These ideal flows show highly curved streamlines, with 
fast jets of plasma emerging from the central region. Although a stronger 
coupling between  the ideal and non-ideal solutions is present in this situation,  we have shown that, just 
as in the non-accelerating case, the ideal flow itself does not directly govern the reconnection rate.
Again the choice of different free functions has been shown to correspond to differing
physical reconnection scenarios;  two particular free functions were used to illustrate how
an acceleration of the plasma may be driven by a Lorentz force or by a pressure gradient. In general, a 
combination of both effects is also possible.

\end{document}